\documentclass[%
 reprint,
 amsmath,amssymb,
 aps,
]{revtex4-2}
\usepackage{graphicx, xcolor}
\usepackage{dcolumn}
\usepackage{bm}
\usepackage{soul}
\usepackage{url}
\usepackage{hyperref}

\begin{document}

\preprint{APS/123-QED}

\title{%Distributed Logical Operations with Enhanced\\Error Suppression in Rotated Surface Codes
Error Suppression in Distributed Quantum Computing\\with Heterogeneous-Distance Lattice Surgery}

\author{Daniel Dilley}
\email{ddilley@anl.gov}
\affiliation{%
Mathematics and Computer Science Division, Argonne National Laboratory, Lemont, IL 60439, USA
}

\author{Anastashia Jebraeilli}
\email{ajebraeilli@anl.gov}
\affiliation{%
Mathematics and Computer Science Division, Argonne National Laboratory, Lemont, IL 60439, USA
}

\author{Rayat Roy}
\email{rayat1@uchicago.edu}
\affiliation{%
Mathematics and Computer Science Division, Argonne National Laboratory, Lemont, IL 60439, USA
}

\author{Shobhit Gupta}
\email{shobhit@memq.tech}
\affiliation{%
memQ Inc., Chicago, IL 60615, USA
}

\author{Alvin Gonzales}
\email{agonzales@anl.gov}
\affiliation{%
Mathematics and Computer Science Division, Argonne National Laboratory, Lemont, IL 60439, USA
}

\author{Zain Saleem}
\email{zsaleem@anl.gov}
\affiliation{%
Mathematics and Computer Science Division, Argonne National Laboratory, Lemont, IL 60439, USA
}

\date{\today}

\begin{abstract}
Distributed quantum computing requires fault-tolerant operations across inter-QPU links that can be substantially noisier than local gates. Uniformly increasing code distance provides additional protection but also enlarges data patches used for local storage and computation. Here, we introduce distributed heterogeneous-distance lattice surgery using an eight-data-patch ancilla-mediated (8-DAM) architecture, which will be useful for near-term quantum devices with less qubit overhead. In this architecture, the central ancilla spanning the inter-QPU boundary is enlarged while the data patches retain distance $d$. The protocol uses traveling stabilizers to suppress hook errors during merge and split operations between these unequal-distance patches. Circuit-level simulations of rotated surface codes at fixed local depolarizing noise show that logical-readout error rates depend only weakly on link noise. The resulting advantage over conventional lattice surgery grows as link errors increase. Comparisons with uniform distance implementations demonstrate comparable logical error suppression with reduced physical-qubit overhead. We also demonstrate how 8-DAM layouts support two simultaneous distributed logical CNOT operations between four logical data qubits using a single enlarged ancilla. At higher link noise, this construction yields lower logical error rates and more stability than two independent distributed logical CNOTs. These results support selective ancilla enlargement as a resource-efficient approach to fault-tolerant distributed quantum computing.
\end{abstract}

\maketitle

\section{\label{sec:intro}Introduction}

Modular quantum architectures offer a route to scaling fault-tolerant computation by distributing logical operations across linked quantum processing units (QPUs)~\cite{Monroe2014, Nickerson2014_topological}. Scaling a single QPU requires managing communication between increasingly distant physical qubits. 
Since unknown quantum states cannot be cloned~\cite{Wooters1982} and quantum information has a finite propagation velocity~\cite{LiebRobinson}, the necessary coupling of distant physical qubits within a single QPU that becomes increasingly larger introduces non-negligible latency and routing overhead~\cite{Bapat2023, Brennen2003}. Simply introducing modularity to quantum infrastructure does not remove the challenge of implementing scalable and reliable quantum operations. Inter-QPU operations are inherently distinct from local gates and are known to introduce higher and communication error rates~\cite{Monroe2014, Nickerson2014_photonic_links} than intra-QPU operations. Reliable logical operations across linked QPU boundaries are essential. This motivates the development of efficient error correction protocols with protection of distributed lattice surgery operations that use these links.

Error-suppression strategies differ in the resources they require and the information they protect. Quantum error mitigation reduces errors in measured observables without explicitly encoding information into a fault-tolerant code, including through circuit cutting and classical post-processing~\cite{Liu2022}. It can also be combined with QEC to suppress residual logical errors after correction~\cite{Zhang2026}. Error-detection approaches identify and discard erroneous outcomes, reducing correction requirements at the cost of post-selection overhead~\cite{Gonzales2025}. Operator quantum error correction provides a framework for protecting encoded subsystems~\cite{Kribs2005}, while correction focuses on preserving task-relevant quantum resources~\cite{Byrd2025}. These approaches involve different tradeoffs between physical-qubit overhead, sampling costs, and the scope of protection.

The surface code is a leading approach to QEC because it has a high error threshold, and geometrically local stabilizer measurements that are innately compatible with two-dimensional qubit layouts~\cite{Bravyi1998, Dennis2002, Fowler2012}. Here, we use the rotated surface code, in which data qubits occupy the vertices of a square lattice with alternating $X$- and $Z$-type stabilizer checks. A particularly important feature of surface codes are their compatibility with lattice surgery as a means of implementing logical operations between discontinuous code patches within a planar layout~\cite{Horsman2012}. A rotated surface code layout also further reduces the physical qubit resource cost per code patch compared to an unrotated surface code at the same distance~\cite{ORourke2025}. A rotated surface code is therefore a natural environment to study the optimal resource allocation for inter-QPU operations.

If the physical error rates fall below a particular error corrections threshold~\cite{Wang2009, Kitaev2003, Bravyi1998, Fowler2012}, increasing the code distance strengthens protection against logical faults. Distributed computation also requires reliable operations between logical qubits on separate QPUs. A transversal surface-code CNOT applies physical CNOT gates between corresponding data qubits in the two patches and has been investigated  experimentally~\cite{Bluvstein2023}. This will break the assumption of local connectivity since it will require longer range connections between the two patches. When the patches occupy separate QPUs, these transversal gates will place even higher demands on inter-QPU connectivity. To the contrary, lattice surgery implements logical operations through joint parity measurements obtained by merging and splitting patch boundaries~\cite{Horsman2012,Chatterjee2025} while maintaining local connectivity. In our work, an ancillary patch is used to mediate the parity measurements that implement a logical controlled-NOT (lCNOT)~\cite{Vuillot2019}. Circuit-level studies have compared the performance of transversal and lattice-surgery CNOTs~\cite{Fang2026}.

Inter-QPU operations have higher error rates and longer latencies than local gates~\cite{Monroe2014,Main2025,Nickerson2014_photonic_links}. Recent studies have shown that distributed lattice surgery can remain fault-tolerant despite this disparity~\cite{Sinclair2023,Chandra2026,Marton2025}. Uniformly increasing patch distances across all quantum processors provides additional protection against logical errors but may require excessive storage space and longer runtime. This motivates a resource-allocation question: how effectively can additional protection be concentrated at the interface between quantum processors while retaining the original code distance of the data qubit patches? Addressing this question is especially relevant for early fault tolerant (EFT) systems where physical qubit budgets remain modest and limited, and inter-QPU operations are unlikely to rival the fidelity or speed of processor-local gates.

We introduce heterogeneous-distance lattice surgery in which the logical data patches retain distance-$d$ and an ancilla spanning the inter-QPU boundary is set to distance $d_A = 2d +1$. The inter-QPU boundary passes through the center of the ancilla patch, while the surgery seams joining it to the data patches lie along its edges; thus, eliminating the need for the data qubits to ever have to directly interact across the boundary. All inter-QPU operations will solely occur in the larger distance ancilla patch, even during merge and split operations with the data qubit patches. This geometry only requires additional qubit overhead around the noisy interface while preserving the original footprint of the logical data patches everywhere else. So the construction targets sensitivity to link errors while the data patches retain their original distance-$d$ protection against localized faults on the processor. We study the control$-$ancilla$-$target configurations $3-7-3$ and $5-11-5$ and compare them with uniform-distance implementations. 

Our method also addresses hook errors, a type of correlated error that arises during syndrome extraction when a fault on a syndrome qubit propagates to two data qubits through two sequential physical CNOT gates ~\cite{Litinski2018, Kishony2026, ORourke2025}. We use a traveling stabilizer construction~\cite{Kishony2026, McEwen2023, Hirai2026}, an approach also discussed under the name ZX interleaving syndrome extraction ($Z \to Z, X \to X$) in prior works~\cite{Hirai2026, McEwen2023}. %This means that our implementation has stabilizer supports that evolve with changing merge and split boundaries during lattice surgery rather than remaining fixed to the original patch geometry.
Traveling stabilizer scheduling prevents hook errors from accumulating into logical fault paths and is essential to preserving fault tolerance when lattice surgery is performed between two surface code patches of different code distances. The logical observables can become directed 90 degrees to their original path, which could cause multiple errors along a logical fault in the standard N/Z scheduling for syndrome extraction.

We evaluate these constructions under circuit-level stochastic depolarizing noise, varying the physical error rates for inter-QPU gates between $p_{\mathrm{link}}=10^{-3}$ and $p_{\mathrm{link}}=10^{-2}$ and fixing all other physical error rates to be $p_{\mathrm{local}}=10^{-4}$. These will include measurement, reset, Hadamard, and CNOT error rates for operations contained on a single QPU. For the single-lCNOT circuits, the decoding failure probabilities exhibit a substantially weaker dependence on link noise than those of the uniform-distance baselines. At $p_{\mathrm{link}}=10^{-2}$, the $3\!-\!7\!-\!3$ construction reduces these failure probabilities by factors of approximately 13 and 19 relative to uniform distance-3. Uniform distance-5 performs better at lower link noise, whereas the heterogeneous construction outperforms it at the upper end of the sweep. At the largest tested link-error probability, the $5\!-\!11\!-\!5$ construction achieves readout failure probabilities comparable to uniform distance-9 while retaining distance-5 data patches.

Our constructions exhibit a weak dependence on link noise in all of our simulations. These metrics characterize logical-observable readout failures per complete circuit rather than total logical-gate failure probabilities. We do not apply any physical corrections since we are only concerned with tracking the logical observable across the inter-QPU boundary. We assume that the local QPU operations will be done fault-tolerantly if the distance is sufficiently high.

The improved tolerance to link noise comes with a resource tradeoff. The $5\!-\!11\!-\!5$ circuit uses 373 distinct physical qubit sites, compared to 521 for uniform distance-9, a reduction of 28.4\%. It also requires less overall inter-QPU operations per QEC cycle, even though it has more distinct inter-QPU couplings. Our model isolates spatially heterogeneous gate errors and excludes idle noise, link-generation failures, heralding latency, leakage, and correlated link noise. In future studies, these hardware constraints can be considered in various settings.

The remainder of this paper is organized as follows. Section~\ref{sec:background} reviews rotated surface codes, distributed lattice surgery, and syndrome extraction scheduling. Section~\ref{sec:protocol} presents the heterogeneous-distance constructions, simulation model, logical-readout results, and resource comparisons. Section~\ref{sec:conclusion} discusses the implications of our work and possible directions of future work.

\section{\label{sec:background}Background}

For readers new to quantum error correction and surface code computation, we recommend the introductory treatments in Refs.~\cite{Fowler2012, Litinski2019, Roffe2019, Chatterjee2025}. Here, we summarize the concepts and definitions needed for our lattice surgery constructions.

\subsection{\label{sec:surface_code_background}Rotated surface-code patches}

A standard square distance-$d$ rotated surface code patch encodes one logical qubit in $d^2$ data qubits~\cite{Horsman2012} and is capable of correcting any Pauli error affecting at most $ \frac{d-1}{2}$ data qubits~\cite{Regev2026} for odd distances. 
The code space is defined by the simultaneous $+1$ eigenspace of commuting $X$- and $Z$-type stabilizer generators. Interior generators act on four neighboring data qubits, while boundary generators act on two. Measurement ancilla repeatedly measure these checks, producing outcomes of either $+1$ or $-1$ that provide information about errors without revealing the encoded logical state~\cite{Fowler2012}.

\begin{figure}
    \centering
    \includegraphics[width=\linewidth]{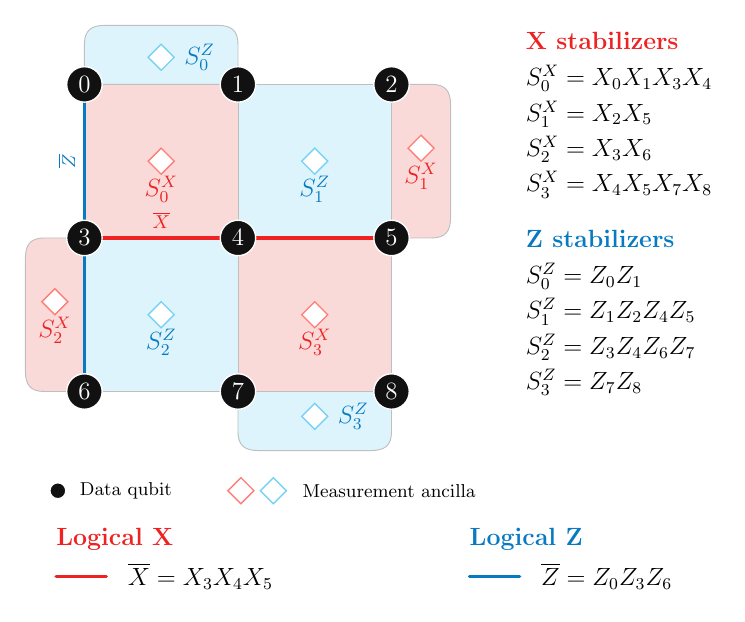}
    \caption{Distance-3 rotated surface code patch and its stabilizer generators. The red horizontal path represents $X_L = \overline{X}$, and the blue vertical path represents $Z_L = \overline{Z}$.}
    \label{fig:logical_obs}
\end{figure}

Figure~\ref{fig:logical_obs} illustrates a distance-3 surface code patch with data qubits numbered $0-8$. Red and blue regions identify the supports of the $X$- and $Z$-type generators, respectively, and the white diamonds denote their respective measurement ancillas. The eight generators are listed beside the patch. For example, the upper-left bulk check is $S^X_0 =X_0 X_1 X_ 3 X_4$, whereas the upper boundary check is $S^Z_0 = Z_0 Z_1$. Here, $X_i \text{ and } Z_i$ denote physical Pauli operators acting on data qubit $i$. An $X$-type check changes sign under an odd number of $Z$ or $Y$ errors on its support, and a $Z$-type check changes sign under an odd number of $X$ or $Y$ errors~\cite{Fowler2012}. The representation of a logical operator is not unique. For example, $X_L S^X_0 S^X_1 = X_0 X_1 X_2$ gives an equivalent representative for a logical $X$ operator along the top edge of data qubits. We label patch edges by the logical operators they support: an $X$ edge supports a representative of $X_L$ running along that edge, and a $Z$ edge similarly supports $Z_L$~\cite{Litinski2019}.

\subsection{\label{sec:lCNOT_background}Logical CNOTs}

Lattice surgery couples logical qubits through joint parity measurements in order to implement a logical controlled-NOT (lCNOT) operation between surface code patches~\cite{Chatterjee2025, Litinski2018, Litinski2019}.  We label logical Pauli operators by their patches. Measuring $Z_C Z_A$ determines the joint $Z$ parity without having to  measure the data  qubits directly; that is, we can extract the parity information for this operator only from syndrome measurements. The $X_A X_T$ measurement similarly determines the joint $X$ parity for the ancilla and target patches during lattice surgery.

To measure a joint parity, stabilizer checks are activated between adjacent patch boundaries to merge the patches and then measured out. When these intermediate qubits are turned on they form the lattice-surgery seam. Products of the appropriate stabilizer outcomes determine the joint logical parity, and the patches are separated by measuring out the seam qubits~\cite{Vuillot2019, Horsman2012}. 

\begin{figure}
    \centering
    \includegraphics[width=\linewidth]{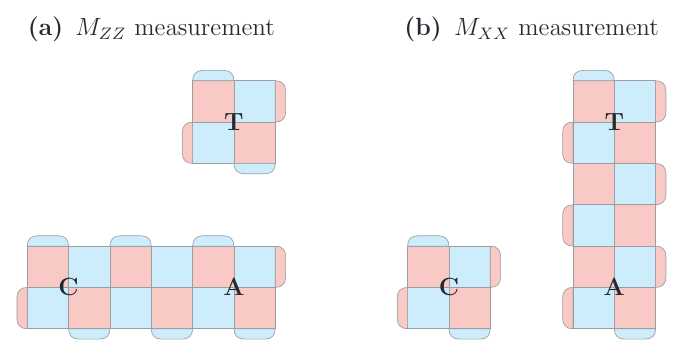}
    \caption{Merge configurations for the two joint parity measurements in a distance-3 lattice surgery lCNOT: \textbf{(a) }$Z_C Z_A$ and \textbf{(b)} $X_A X_T$. Red and blue plaquettes denote $X$- and $Z$-type stabilizers, respectively.}
    \label{Fig:Standard_lCNOT}
\end{figure}

One standard lCNOT using control, ancilla, and target patches $C$, $A$, and $T$, respectively, prepares the ancilla patch $A$ initially in $|+ \rangle_L$, measures $Z_C Z_A$, then measures $X_A X_T$, and finally measures out the ancilla patch $Z_A$. The measurement outcomes determine any Pauli corrections needed for the control and target patches and can be tracked classically in a Pauli frame, which specifies how subsequent operations and measurements should be interpreted~\cite{Litinski2018}. Figure~\ref{Fig:Standard_lCNOT} shows the two merge configurations used for these joint parity measurements; the patches must undergo $d$ QEC cycles during each merge step. %The parity order, ancilla preparation, and final ancilla readout must be specified together for each implementation.
Lattice surgery can also join patches of unequal sizes~\cite{Litinski2019, Horsman2012}. Our constructions in Section~\ref{sec:protocol} use $d_C = d_T = d$, and $d_A = 2d + 1$ to concentrate additional code resources in the interface ancilla while retaining smaller data patches. We must use $2d-1$ QEC cycles to maintain fault tolerance with the larger ancilla patch.

\subsection{\label{sec:distributed_qc_background}Distributed and modular fault tolerant quantum computing}

Modular, distributed quantum architectures enable scaling beyond the limits of a monolithic QPU by supporting remote interactions between code patches on separate QPUs. These interactions are mediated by inter-QPU quantum interconnects built on quantum channels such as optical fibers or microwave transmission lines \cite{Sinclair2023,Marton2025}. Schemes for inter-QPU interconnects broadly fall into two categories: deterministic schemes and probabilistic heralded schemes \cite{teoh2025robustquantumcommunicationlossy,mohseni2026buildquantumsupercomputerscaling}. Deterministic schemes use the quantum link as a communication bus that mediates direct quantum state transfer, mapping a qubit state onto a flying mode such as a photon \cite{Axline2018-by,Kurpiers2018-rh}. Heralded schemes instead first generate Bell pairs across the link and then consume them probabilistically for inter-module state and gate teleportation \cite{Sinclair2023}. Across both schemes, typical Bell-state fidelities lie between 93\% and 97\% for optical links between trapped-ion qubits \cite{Saha2025-nu,StephensonPhysRevLett.124.110501,OReillyPhysRevLett.133.090802}, and between 85\% and 94\% for microwave links between superconducting qubits \cite{nv7d-k3wr,yanPhysRevLett.128.080504,PRXQuantum.2.030321}.

Bell state fidelities serve as a proxy for inter-module two-qubit gate fidelity, and corresponding error rates, roughly 3\% to 15\%, exceed the local two-qubit gate error rates of $5 \times 10^{-4}$ for ions \cite{hughes2025trappediontwoqubitgates9999} and $1.4 \times 10^{-3}$ for superconducting qubits \cite{blogMeetWillow} by roughly one to two orders of magnitude. Despite this gap, recent work has shown that the surface code exhibits a threshold for inter-module errors roughly an order of magnitude higher than the local-operation threshold, which makes present-day noisy links useful for near-term fault-tolerant computation without the added overhead of entanglement distillation \cite{Ramette2024-to, Sinclair2023}. A recent experimental demonstration of distributed error correction across trapped-ion modules connected by noisy photonic links further argues that a fault-tolerant distributed architecture can tolerate substantially higher inter-QPU error rates, provided the local gates remain sufficiently high fidelity \cite{ainley2026errorcorrectiondistributedquantum}. We therefore model the inter-QPU interface with higher error rates than intra-QPU bulk.

The inter-QPU boundary defines the physical separation between quantum processors and determines which gates use the link. The surgery seam is the region where logical patches are joined. These regions may coincide, but need not do so. In our circuit constructions, the enlarged ancillas span the physical boundary, while surgery seams with data patches lie along the edges.

\begin{figure}
    \centering
    \includegraphics[width=\linewidth]{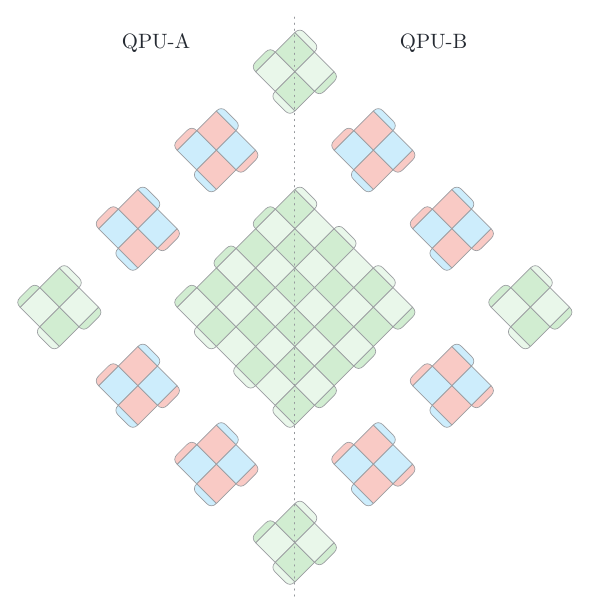}
    \caption{8-DAM patch architecture rotated by $45^\circ$. The vertical gray dotted line marks the inter-QPU boundary separating QPU-A and QPU-B.}
    \label{fig:8dam_rot}
\end{figure}

A noisy interface is also known to affect the two logical observables differently~\cite{Sinclair2023,Chandra2026}. We use circuits with geometries such that the logical $X$ observable is defined perpendicularly to the inter-QPU boundary. A logical $Z$ error string running along the interface can flip this observable. A spatial logical error string that flips the parallel logical $Z$ observable must also extend through the bulk. Our architecture's geometry has been chosen in this work such that the $X$ observable is more vulnerable even when the physical noise is unbiased. This stems from the noisier inter-QPU boundary being aligned with a logical $Z$ operator. However, a choice of an interface orientation that is perpendicular to ours would result in a $Z$ observable that is more vulnerable. One can even rotate by $45^\circ$ to make this unbiased for the logical observables and have the operations only connecting data qubits that are perfectly aligned across the boundary (see Fig.~\ref{fig:8dam_rot}).

\subsection{\label{sec:hook_errors_traveling_stabs_background}Hook errors and traveling stabilizers}
Quantum states are usually prone to physical error rates on the scale of $10^{-4}$ to $10^{-1}$ rendering them useless for accurate quantum computation at utility-scale. The answer to this problem is a quantum error correction code that uses multiple stabilizer measurements over many rounds to ensure that errors are caught by a classical decoder and corrected. Conventional logical CNOT gates (see Fig. \ref{Fig:Standard_lCNOT}) with uniform patch distance-$d$ use $d$ rounds of syndrome measurements to measure a joint parity at distance $d$~\cite{Fowler2012, Vuillot2019}. For our heterogeneous-distance geometry, the required repetition depends on the size of the larger ancilla patch that mediates the lattice surgery for the logical CNOT gates. 

\begin{figure}
    \centering
    \includegraphics[width=\linewidth]{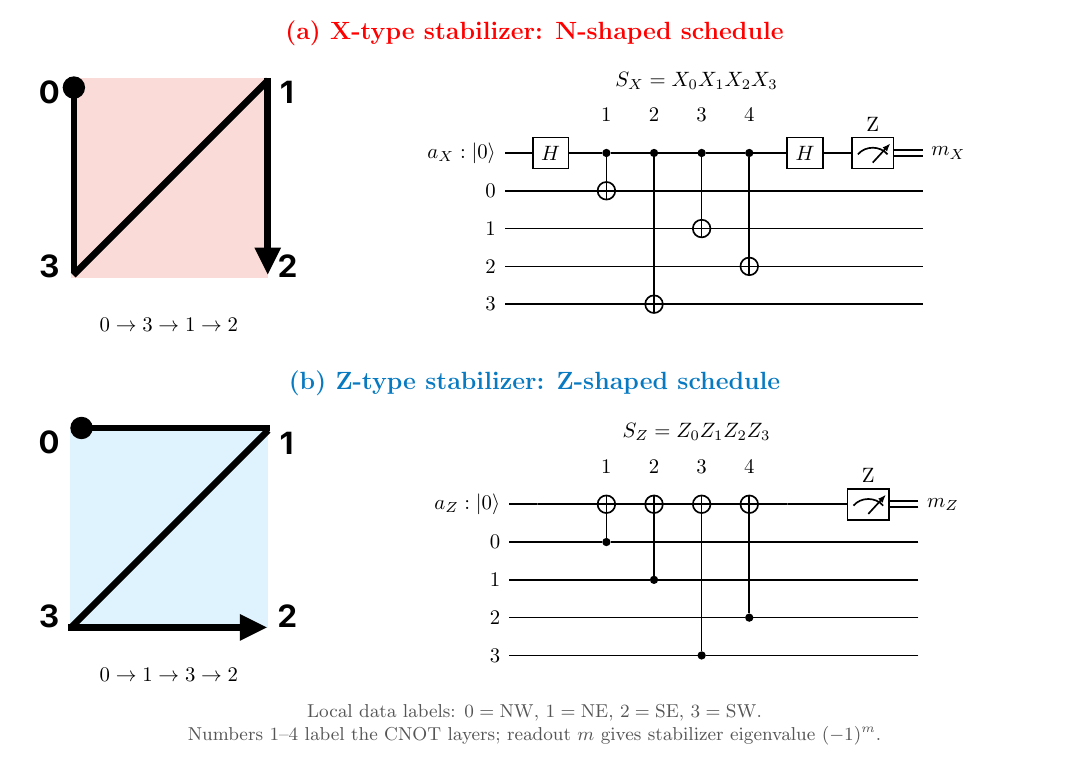}
    \caption{Conventional CNOT stabilizer scheduling and their corresponding measurement circuits in a rotated surface code. \textbf{(a)} The $X$-type stabilizers (red, top) follow an ``N" shaped order $0 \to 3 \to 1 \to 2$, with the syndrome qubit controlling each CNOT. \textbf{(b)} The $Z$-type stabilizer (blue, bottom) follows the ``Z" shaped order $0 \rightarrow 1 \to 3 \to 2$, with syndrome qubits as the CNOT target.}
    \label{fig:nz_polygons}
\end{figure}

For our distributed lattice surgery protocol, the order of the CNOT gates used for syndrome extraction becomes important, since merged patches cause the logical observables to curve. A single fault on a syndrome qubit can propagate through later CNOTs and produce multiple data qubit errors. If these errors align with a logical observable of the same time it is called a hook error~\cite{Dennis2002}, which causes the effective distance of the logical patch to decrease. For odd numbered code distance $(d-1)/2$ physical errors are correctable. The conventional $N/Z$ schedule shown in Fig.~\ref{fig:nz_polygons} is all that is necessary to avoid hook errors in the standard logical CNOT gates during merge operations when patches have the same code distance.

\begin{figure*}
    \centering
    \includegraphics[width=\linewidth]{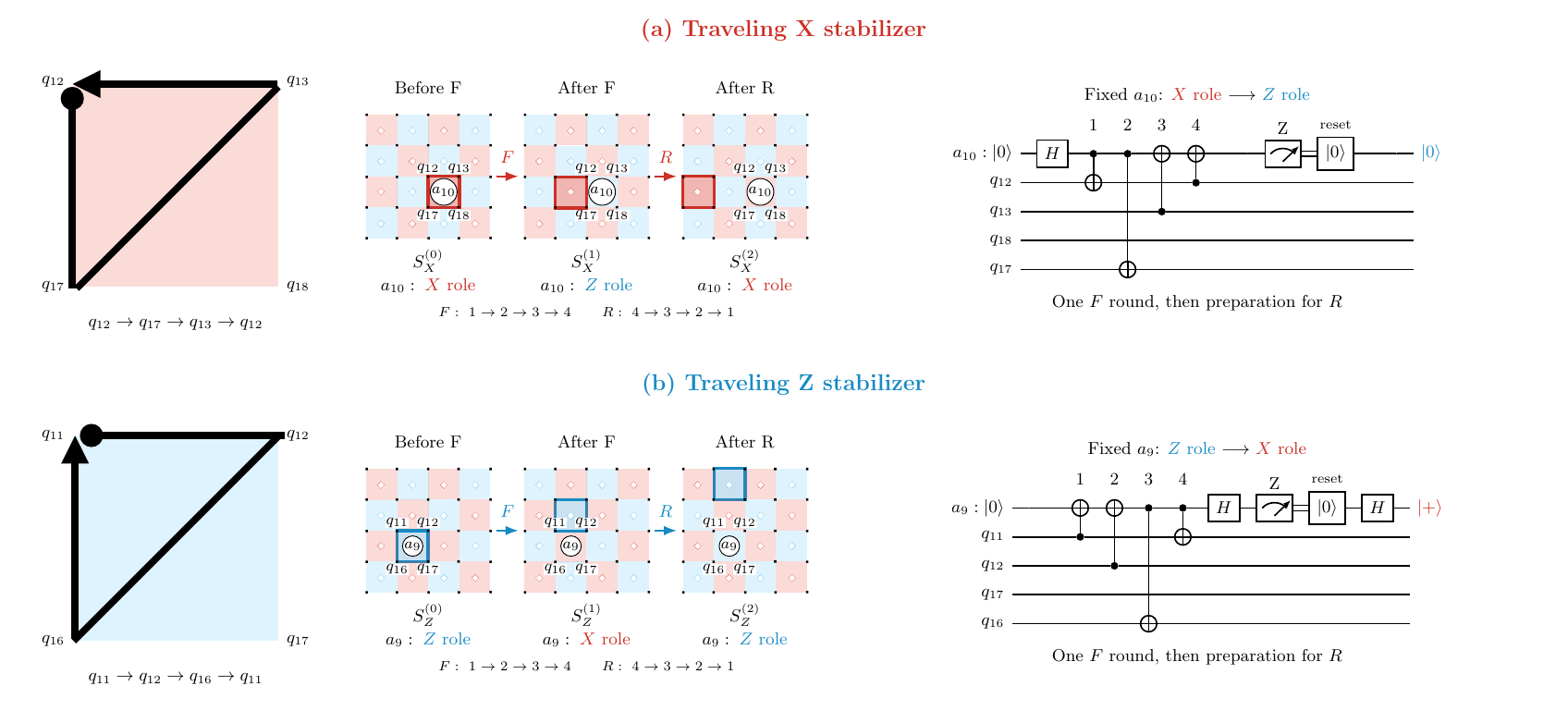}
    \caption{Traveling stabilizer construction for \textbf{(a)} $X$-type (red, top) and \textbf{(b)} $Z$-type (blue, bottom) checks. Black circles labeled $q_j$ represent data qubits; $a_j$ represent syndrome qubits. The left panels show the local CNOT interaction order during a forward (F) round, with physical-qubit labels matching the checkerboard snapshots and circuits. The central panels show the successive supports $S_P^{(0)}$, $S_P^{(1)}$, and $S_P^{(2)}$, with $P\in\{X,Z\}$, before F, after F, and after a reverse (R) round. The F and R rounds apply the four CNOT layers in the orders $1,2,3,4$ and $4,3,2,1$, respectively. The highlighted $X$ supports move leftward and the $Z$ supports move upward as the checkerboard alternates, while all physical qubits remain stationary. The right panels show the four CNOTs involving the selected syndrome site during F, followed by readout and preparation for R. The fixed site $a_{10}$ changes from $X$ preparation ($\lvert+\rangle$) to $Z$ preparation ($\lvert0\rangle$), while $a_9$ undergoes the complementary change. These local circuit excerpts omit neighboring-site interactions required for the complete traveling stabilizer construction.}
    \label{fig:ts_polygons}
\end{figure*}

Traveling stabilizers~\cite{Hirai2026, McEwen2023} are a clever way to avoid hook errors during the merge of patches in all the settings we consider. One example of how they can work is that the $Z$ stabilizer can travel upward while the $X$ stabilizer travels to the left. So the $Z$ stabilizer plaquettes begin with two CNOT gates in the Z schedule and then end with two CNOT gates in the $X$ schedule, and vice versa. Figure~\ref{fig:ts_polygons} illustrates what the schedule of the stabilizers look like everywhere in the interior of the rotated surface code patches. One unapparent advantage of the traveling stabilizers is that you can choose the direction they move to minimize the number of inter-QPU operations across the boundary during a QEC cycle, while also reducing the number of QEC cycles by 1 at any distance. When splitting a logical patch of distance-$d$, it is always possible to ensure only a total of $d$ inter-QPU operations are required across the inter-QPU boundary for an entire QEC cycle. This savings becomes noticeable over the standard N/Z schedule, which requires $2d-1$ inter-QPU operations, when the operations between quantum processors are noisy.

Scheduling choices throughout the circuit's successive gate layers also affects the physical interactions required at the QPU boundary. A different traveling construction can change the total number of inter-QPU operations needed during lattice surgery. By using the incorrect construction, it is possible to require more inter-QPU operations than in the N/Z schedule. 

Lattice surgery can join patches of unequal size~\cite{Horsman2012, Litinski2019} as long as the logical space is preserved and the number of syndrome qubits stays consistent with the number of data qubits during the merge and split operations. Notice that the distance-3 logical qubit in Fig. (\ref{fig:logical_obs}) follows the $[n, k, d]$ notation of Ref.~\cite{Calderbank1997} with $n = 9, k = 1$, and $d = 3$. Since there are 9 data qubits and 8 stabilizers, we are encoding 1 logical qubit in the rotated surface code patch, which must also be true when merging and splitting for a single lCNOT gate. This is different for the dual distributed lCNOT gates since the ancilla patch is now encoding 2 logical qubits, which means we should have two more data qubits than syndrome qubits during the merge.

By joining unequal patch sizes, lattice surgery allows for the control, ancilla, and target patches, denoted as $d_C - d_A - d_T$, to be chosen separately. In this work we define our patch geometry such that $d_C = d_T = d$, and $d_A = 2d + 1$. The data patches retain their chosen distance-$d$ for local operations, and the larger ancilla patch provides additional resources at the inter-QPU interface. 

\subsection{\label{DEM_background}Detector error models and decoding}

A detector is a parity of measurement outcomes whose value is deterministic in an equivalent but noiseless circuit. A detection event occurs when this parity differs from its expected value. For repeated measurements of an unchanged stabilizer, a detector can compare consecutive outcomes; detectors near initialization, readout, and merge or split boundaries must follow the corresponding circuit relations. Therefore, a detector error model is necessary to run QEC simulations that track when logical faults occur from physical errors that cause detection events. This representation connects the syndrome-extraction circuit to decoding and is supported by Stim~\cite{Gidney2021}.

A limitation of Stim is that only Clifford circuits can be simulated efficiently on classical hardware. Our simulations also do not account for additional sources of noise, including non-Clifford noise, idling errors, and correlated noise, which would be expected to further increase the observed error rates. However, because our analysis compares the evaluated approaches under the same noise assumptions, we expect the relative comparisons to remain valid when these additional noise sources are incorporated consistently across all experiments.

For matching-based decoding, graph-like error components connect two detector vertices, or one vertex to a decoding boundary. Edge weights encode their relative likelihoods. We use minimum weight perfect matching (MWPM) implemented in PyMatching~\cite{Higgott2021}, with correlated matching disabled. The decoder finds a minimum-weight edge configuration consistent with the observed detection events. This optimization uses the matching model's assumptions and does not generally provide maximum-likelihood decoding of the full circuit noise model. 

%\textbf{Successful readout decoding requires correctly predicting the measured logical-observable flips without reconstructing every physical fault. For a static code, corrections differing by a stabilizer have the same logical action. In our circuit experiments detailed in Section~\ref{sec:protocol}, we estimate failure probabilities by comparing predicted and sampled observable flips over complete circuits. The control-$X$ and target-$Z$ results characterize the corresponding readouts rather than the total logical-gate failure probability. }

In our simulations, the decoder uses detection events to predict the flip of a tracked logical observable. We compute the readout failure probability as the fraction of completed circuit executions that disagree with the prediction sampled by Stim,
\begin{equation}
    \widehat{p}_{\mathrm{L}} =
    \frac{1}{N}\sum_{i=1}^{N} e_i, \qquad e_i = 1- \delta_{\ell_i, \widehat{\ell}_i},
\end{equation} where $N$ is the number of simulated shots, $\ell_i \in \{0,1\}$ is the logical observable flip samples by Stim, and $\widehat{\ell}_i$ is the decoder's prediction for that shot. The Kronecker delta, 
\begin{equation}
    \delta_{\alpha, \beta} = 
    \begin{cases}
        1, & \text{if } \alpha = \beta, \\
        0, &\text{if } \alpha \neq \beta,
    \end{cases}
\end{equation}so $e_i$ is one for an incorrect prediction and zero when the prediction is faithful. The results presented in Section~\ref{sec:protocol} characterize these separate readout experiments, not the total logical gate failure probability.

\section{\label{sec:protocol}Protocol and Results}

Here we present our methodology to investigate depolarizing noisy inter-QPU operations on lCNOT operations. We implement these logical gates using lattice surgery between rotated surface code patches without using the same code distance for the patches being merged. Our constructions use a larger ancillary code patch with the inter-QPU boundary placed in its center for better logical suppression across the boundary. The data qubit patches never interact directly with the inter-QPU boundary. %We present the heterogeneous-distance construction, compare its logical-readout performance and resource requirements with uniform-distance baselines, and extend it to simultaneous lCNOTs mediated by a shared ancilla.

\subsection{\label{sec:architectures_and_their_diagrams}Heterogeneous distance architecture}

\begin{figure}[h!]
    \centering
    \includegraphics[width=\linewidth]{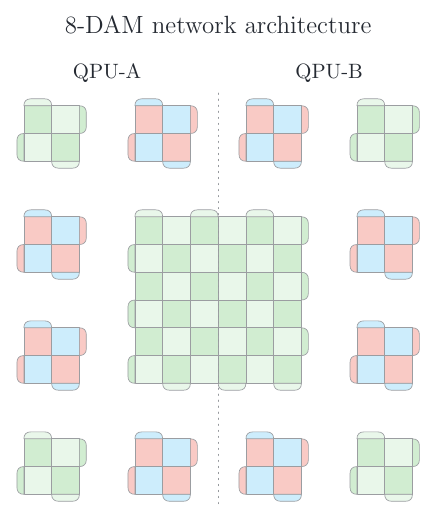}
    \caption{The 8-DAM network for distance-3 logical data-qubit patches, featuring a $7 \times 7$ central ancilla patch that mediates logical operations between data qubits. Green patches denote ancillary logical qubits. The inter-QPU boundary is defined by the gray line separating the right and left portions of the network.}
    \label{Fig:8_DAM_Network}
\end{figure}

Figure~\ref{Fig:8_DAM_Network} illustrates the 8-data-patch ancilla-mediated (8-DAM) geometry for $d = 3$. A central $7 \times 7$ ancillary code patch provides access to eight possible neighboring data code patch positions distributed across two QPUs denoted as QPU-A and QPU-B. The inter-QPU boundary is deliberately chosen to pass centrally through the distance-7 ancillary patch. Qubits on that boundary are contained in QPU A, and any gate that acts across the boundary to those qubits will have their own noise model.

\begin{figure*}
    \centering
    \includegraphics[width=\linewidth]{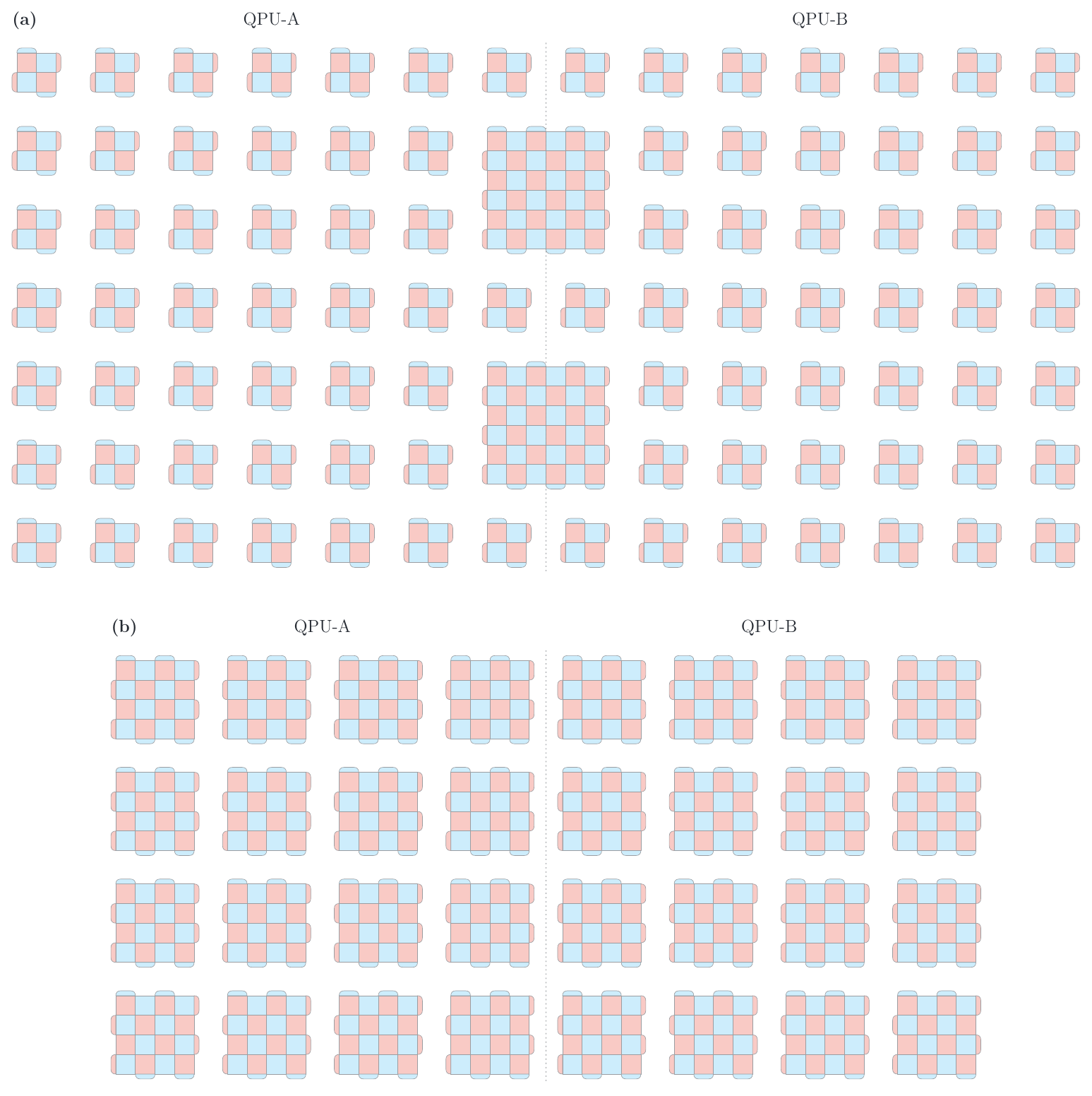}
    \caption{Using the 8-DAM network, logical error rates can be maintained across the inter-QPU boundary without having to increase the distance of the data qubit and ancilla patches that are not placed across the boundary. \textbf{(a)} A total of up to 49 distance-3 patches can be positioned on each QPU. \textbf{(b)} Increasing the distance of all the patches allows for only 16 logical qubits on each quantum processor.}
    \label{fig:standard_alternate_architecture}
\end{figure*}

Figure~\ref{fig:standard_alternate_architecture} illustrates the modular design that includes our 8-DAM construction and shows a key motivation for retaining smaller code patches away from the interface. By enlarging boundary ancillas, more physical qubits are available for additional logical encoding and operation.

\subsection{\label{sec:schedules_and_protocol}Distributed lCNOTs and their syndrome scheduling}

In addition to using the \hyperref[Fig:8_DAM_Network]{8-DAM network architecture} introduced in Subsection~\ref{sec:architectures_and_their_diagrams} we also implement a syndrome schedule that uses traveling stabilizers whose support follows the moving stabilizer plaquettes to avoid hook errors. 

\subsubsection{\label{sec:one_lcnot_protocol}lCNOT geometry}

\begin{figure}[h!]
    \centering
    \includegraphics[width=1\linewidth]{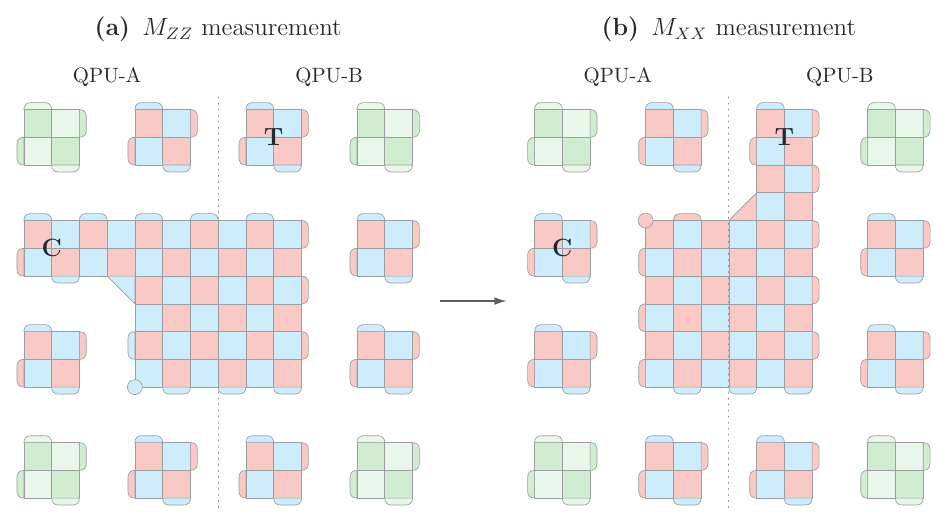}
    \caption{A single distributed lCNOT gate between data-qubit patches C (control) and T (target). First, an \textbf{(a)} $M_{ZZ}$ measurement is performed between the central ancilla patch and C, followed by an \textbf{(b) }$M_{XX}$ measurement between T and the central ancilla patch.}
    \label{Fig:Single_lCNOT}
\end{figure}

Figure~\ref{Fig:Single_lCNOT} depicts a single distributed lCNOT between control patch $C$ and target patch $T$. We are specifically tracking the logical $Z$ observable for these simulations. The central ancilla patch $A$ is prepared in $|+\rangle_L$, followed by measurements of $Z_C Z_A$ and $X_A X_T$, and a final $Z_A$ readout. The measurement outcomes determine the Pauli-frame updates used to interpret the outputs~\cite{Vuillot2019}. However, these are not necessary for our simulation results since we are only concerned with tracking logical errors for the observable that crosses the inter-QPU boundary. The small red and blue circles at the corner of the ancilla patch are necessary for the traveling stabilizer (ZX interleaving schedule with $Z \rightarrow Z$ and $X \rightarrow X$) formalism.

Our simulation comparisons presented in Subsection~\ref{sec:single_lcnot_data} are for tracking the logical X observable. In this case, we run the circuit in Figure~\ref{Fig:Single_lCNOT} in reverse; that is, we perform the $X_A X_T$ first and then the $Z_C Z_A$ measurement right after. The ancilla patch is initiated in the $|0\rangle$ state and measured out in the $X$ basis, but the data qubit patches are now initiated and measured out in the $X$ basis. The control and target will be switched as well. The $3-7-3$ target-$Z$ circuit and the $5-11-5$ circuits use the same conventions. 

\subsubsection{\label{sec:dual_diagram}Dual lCNOT geometry and routing}

\begin{figure}
    \centering
    \includegraphics[width=1\linewidth]{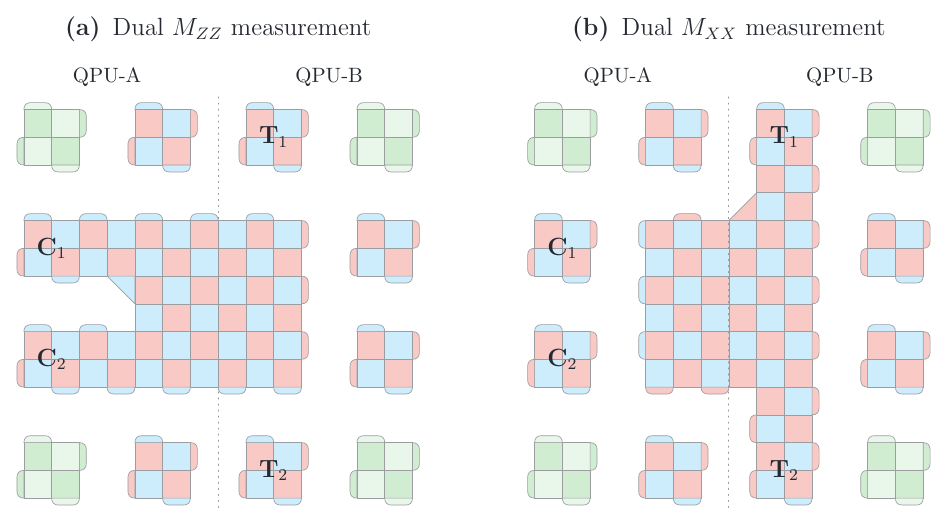}
    \caption{Two distributed lCNOT gates are implemented simultaneously using a single larger central ancilla patch, providing increased error suppression. The top and bottom control--target pairs are $(C_1,T_1)$ and $(C_2,T_2)$, respectively. \textbf{(a) }Two $M_{ZZ}$ measurements are made in step 1 and \textbf{(b)} two $M_{XX}$ measurements are made in step 2.}
    \label{Fig:Dual_lCNOTs}
\end{figure}

Figure~\ref{Fig:Dual_lCNOTs} shows a possible routing protocol using 8-DAM geometry for four distance-$3$ patches corresponding to lCNOT pairs $(C_1, T_1)$ and $(C_2, T_2)$ to implement dual simultaneous lCNOTs using a single larger central ancilla patch. This is possible when the ancilla patch encodes two logical qubits by having 6 edges \cite{Litinski2019} (3 $Z$-type and 3 $X$-type) instead of only 4 (2 $Z$-type and 2$X$-type), like in the single lCNOT scenario. The schedule evaluated in this work uses six syndrome extraction rounds during both merge operation prior to splitting. This ensures fault tolerance is maintained during lattice surgery.

We checked that the logical CNOT gates were properly implemented with that single ancilla patch while mediating both lCNOT gates by tracking the control and target states of the data qubit patches. Over all 16 possible initial state configurations for the four data patches, we always obtained the correct output logical states after our protocol was implemented. This happens because the blue edge that curves between the control patches in Fig. \ref{Fig:Dual_lCNOTs} acts as the $Z_{C_1} Z_{C_2}$ Pauli operator, while the top (bottom) blue edge acts as the $Z_{C_1}$ ($Z_{C_2}$) Pauli operator. So the logical operators must be defined using a curved path, which requires the traveling stabilizers to avoid hook errors. The results for tracking the logical $Z$ observable errors are presented in Subsection~\ref{sec:dual_lcnot_data}.

\subsubsection{\label{sec:nnn_rounting}Nearest neighbor routing}

\begin{figure}[h!]
    \centering
    \includegraphics[width=1\linewidth]{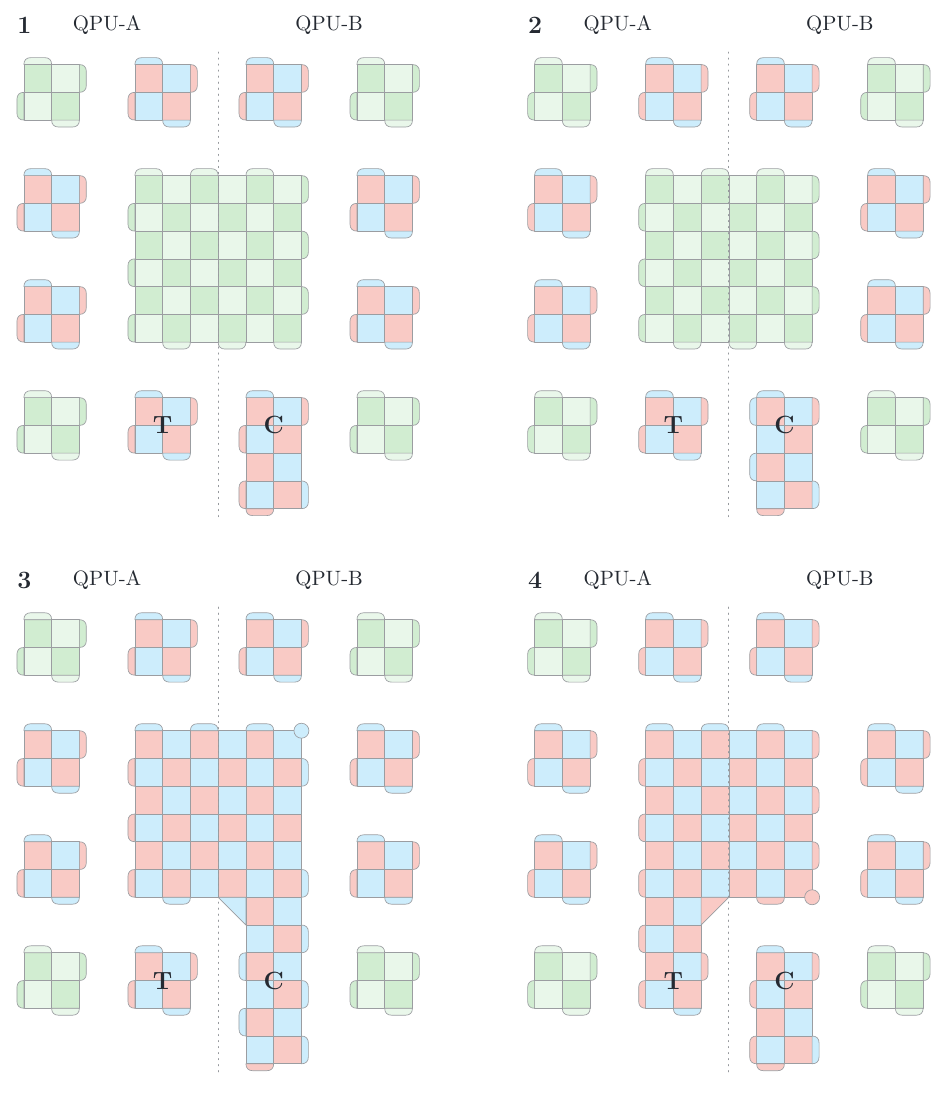}
    \caption{Steps for performing a nearest-neighbor distributed lCNOT gate. First, the control data-qubit patch is grown and deformed prior to the $M_{XX}$ measurement. Next, the control data-qubit patch is further deformed while the target data-qubit patch undergoes an $M_{ZZ}$ measurement. The final step is not shown, as it consists only of measurements that return the control patch to its original position and orientation without requiring additional rounds of stabilization.}
    \label{Fig:Nearest_Neighbor_lCNOT}
\end{figure}

Figure~\ref{Fig:Nearest_Neighbor_lCNOT} illustrates an implementation of a nearest neighbor lCNOT operation across quantum processors using our 8-DAM geometry. The control patch is grown and its edges are rotated to give access to the required joint parity measurements that mediate the logical CNOT gate. The final split between the ancilla and target patches is not shown, but this is also when the control patch is shrunk and returned to its original position and orientation. We include Figure~\ref{Fig:Nearest_Neighbor_lCNOT} as an additional routing example to emphasize the geometric flexibility of 8-DAM. 

The use of the same underlying lattice surgery primitives in the routing of Figure~\ref{Fig:Nearest_Neighbor_lCNOT} and Figures~\ref{Fig:Single_lCNOT} and~\ref{Fig:Dual_lCNOTs} suggests that a comparable logical-error performance may be achievable for the fault-tolerant nearest neighbor routing deformation schedule. The numerical results of this work are restricted to
the explicitly simulated routing constructions, and so the testing of this expectation for the nearest neighbor arrangement remains to be quantified in future work. 

\subsection{\label{sec:sim_model_protocol}Simulation model}
We evaluate the 8-DAM based single-lCNOT and dual-target schedules introduced in Subsections~\ref{sec:architectures_and_their_diagrams} using circuit-level simulations in Stim~\cite{Gidney2021, Fowler2012} and Sinter~\cite{sinter}. The local depolarizing parameter is fixed at $p_{\text{local}} = 10^{-4}$. Single depolarizing noise is applied after single-qubit gates and resets and two-qubit depolarizing noise is applied after local CNOTs. Measurement outcomes are flipped with probability $p_{\text{local}}$. For each circuit, the inter-QPU boundary is held fixed and CNOTs crossing that boundary receive a two-qubit depolarizing parameter 10-100 times the local parameter such that $p_{\text{link}} = 0.001, 0.002, ..., 0.01$. The local operations retain the same noise level $p_{\text{local}}$ throughout the sweep. This effective gate-noise model isolates the effect of heterogeneous operation errors without including idle noise, link-generation failures, heralding latency, or leakage. We specifically simulate logical CNOT gates; however, similar results should be attainable for other merge/split scenarios like the ones seen here \cite{Litinski2019}.

We used Crumble~\cite{GidneyCrumble} for interactive
circuit editing and inspection, including tracking logical Pauli
observables through the merge and split stages. The mixed-distance circuits use the traveling stabilizer schedules described in Subsection~\ref{sec:hook_errors_traveling_stabs_background} are evaluated with these Stim-based simulations and decoded with standard (correlated matching disabled) PyMatching~\cite{Higgott2021}. The reported logical error rates are decoding failure probabilities per complete circuit for the control-$X$ or target-$Z$ observable. Separate experiments test the logical relations \[X_C^{\text{in}}X_T^{\text{in}} \rightarrow X_C^{\text{out}}, \quad Z_C^{\text{in}}Z_T^{\text{in}} \rightarrow Z_T^{\text{out}}, \]
respectively, with the measurement-dependent Pauli-frame updates included in the observable definitions. These are observable-specific readout metrics rather than
a total logical-gate failure probability. No shots are discarded.

For the dual-lCNOT target-\(Z\) results, 45 million complete
dual-circuit shots were simulated for each value of
\(p_{\mathrm{link}}\). The top-target, bottom-target, and
either-target failure probabilities are estimated from the
same set of shots, with pointwise \(95\%\) Wilson confidence
intervals. For the independent-pair reference, the single-circuit interval endpoints are transformed using the same probability relation as the central estimate. 

All simulated circuits used in this work will be made publicly available on \hyperlink{https://github.com/aaj10789/8DAM/}{GitHub}~\cite{Jebraeilli_Dilly}.

\subsection{\label{sec:single_lcnot_data}Heterogeneous-distance single lCNOT performance}

We first examine whether enlarging the ancilla can suppress sensitivity to link noise while retaining the original data-patch distances. We label each single-lCNOT circuit by its control, ancilla, and target distances, \(d_C\! -\! d_A\! -\! d_T\). Using our mixed distance architecture, only the ancilla is enlarged. Our method preserves the size of the data patches while placing additional protection only along the noisy interface. In the 8-DAM constructions introduced in Subsection~\ref{sec:architectures_and_their_diagrams}, only the ancilla is enlarged, concentrating the additional code resources at the inter-QPU interface.

\begin{figure}
    \centering
    \includegraphics[width=\linewidth]{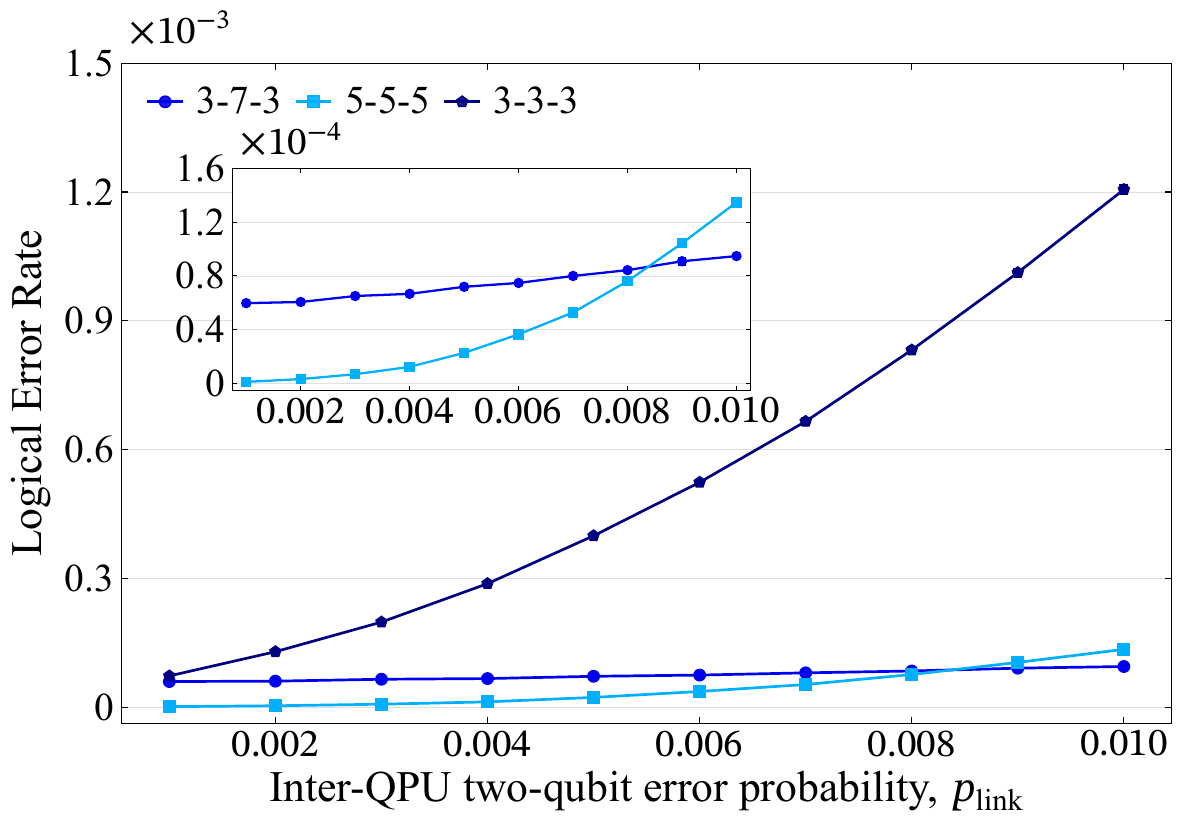}
    \caption{Control-\(X\) decoding failure probability per CNOT for mixed $3-7-3$ (blue circles), uniform distance-5 (cyan squares), and uniform distance-3 (navy pentagons) circuits. Distances are listed as control--ancilla--target. The local noise parameter is \(10^{-4}\), while \(p_{\mathrm{link}}\) is varied. Each point uses 45 million shots.} %The mixed circuit is less sensitive to link noise.}

    \label{fig:mixed_small_x}
\end{figure}

\begin{figure}
    \centering
    \includegraphics[width=\linewidth]{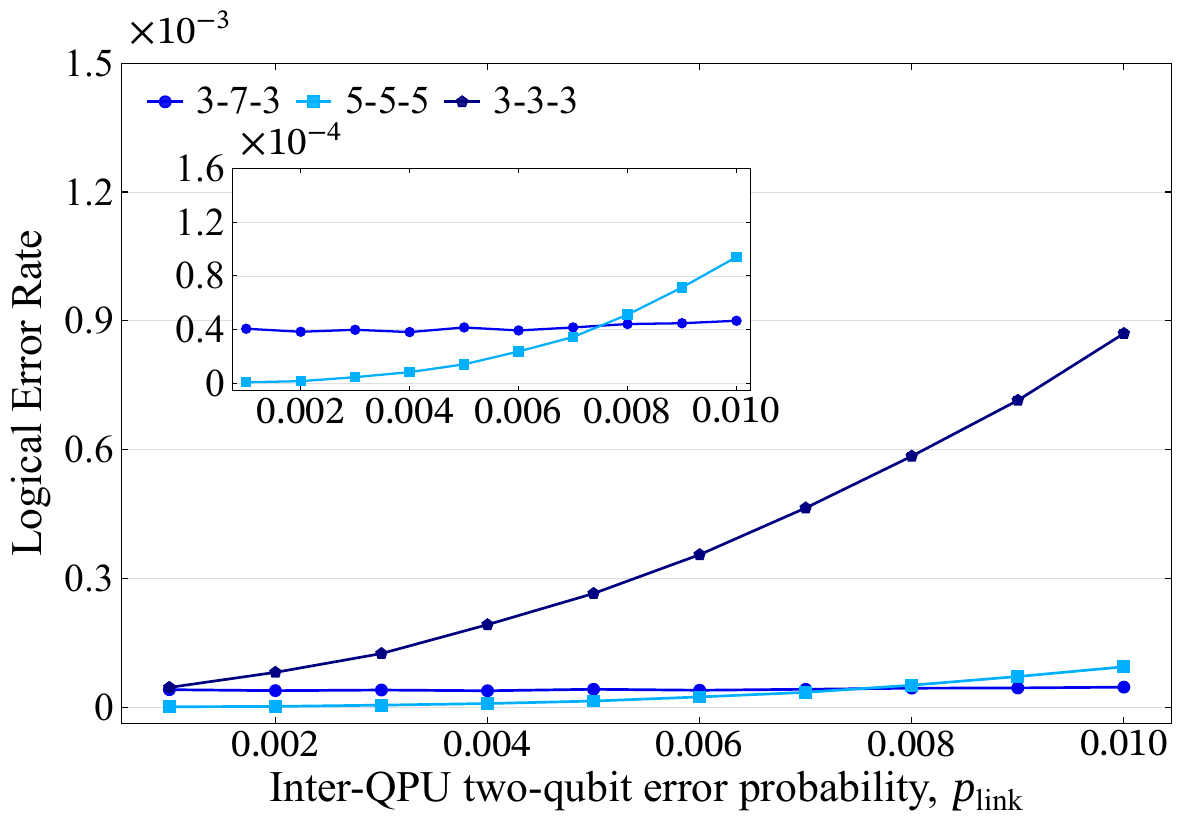}
    \caption{Target-\(Z\) decoding failure probability per CNOT for mixed $3-7-3$ (blue circles), uniform distance-5 (cyan squares), and uniform distance-3 (navy pentagons) circuits, with local noise parameter \(10^{-4}\). Each point uses 45 million shots.} %Our mixed circuit is nearly insensitive to link noise over this sweep and outperforms both uniform circuits at the upper end of the tested error rates.}

    \label{fig:mixed_small_z}
\end{figure}

Figures~\ref{fig:mixed_small_x} and~\ref{fig:mixed_small_z}
 show a comparison of our $3-7-3$ mixed architecture circuit with uniform distance-3 and distance-5 circuits. Both measured failure probabilities depend much less strongly on link noise in our mixed architecture circuit than in the uniform circuits. At \(p_{\mathrm{link}}=0.01\), the mixed circuit gives control-\(X\) and target-\(Z\) failure probabilities of \(9.47\times10^{-5}\) and \(4.67\times10^{-5}\), respectively: approximately 13 and 19 times lower than the uniform distance-3 results. Uniform distance-5 performs better at low link noise, but the mixed circuit performs better and with a greater stability for both observables with increasing noise levels. 

\begin{figure}
    \centering
    \includegraphics[width=\linewidth]{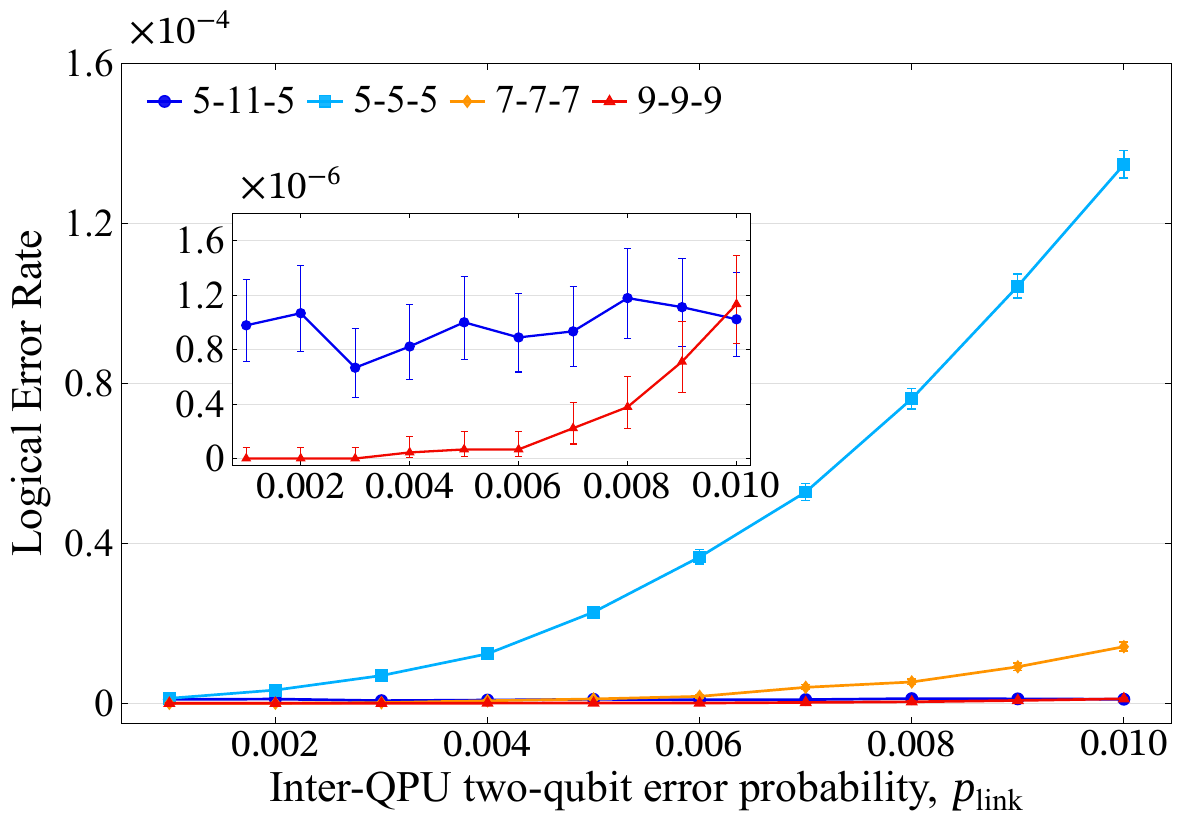}
    \caption{Control-\(X\) decoding failure probability per CNOT for mixed $5-11-5$ (blue circles) and uniform distance-5, -7, and -9 circuits (cyan squares, orange diamonds, and red triangles), with local noise parameter \(10^{-4}\). The inset resolves the mixed and distance-9 results. Each point uses 45 million shots; error bars show 95\% Clopper--Pearson intervals.} %The mixed and distance-9 results are comparable at the largest tested link-noise probability.}

    \label{fig:mixed_large_x}
\end{figure}

\begin{figure}
    \centering
    \includegraphics[width=\linewidth]{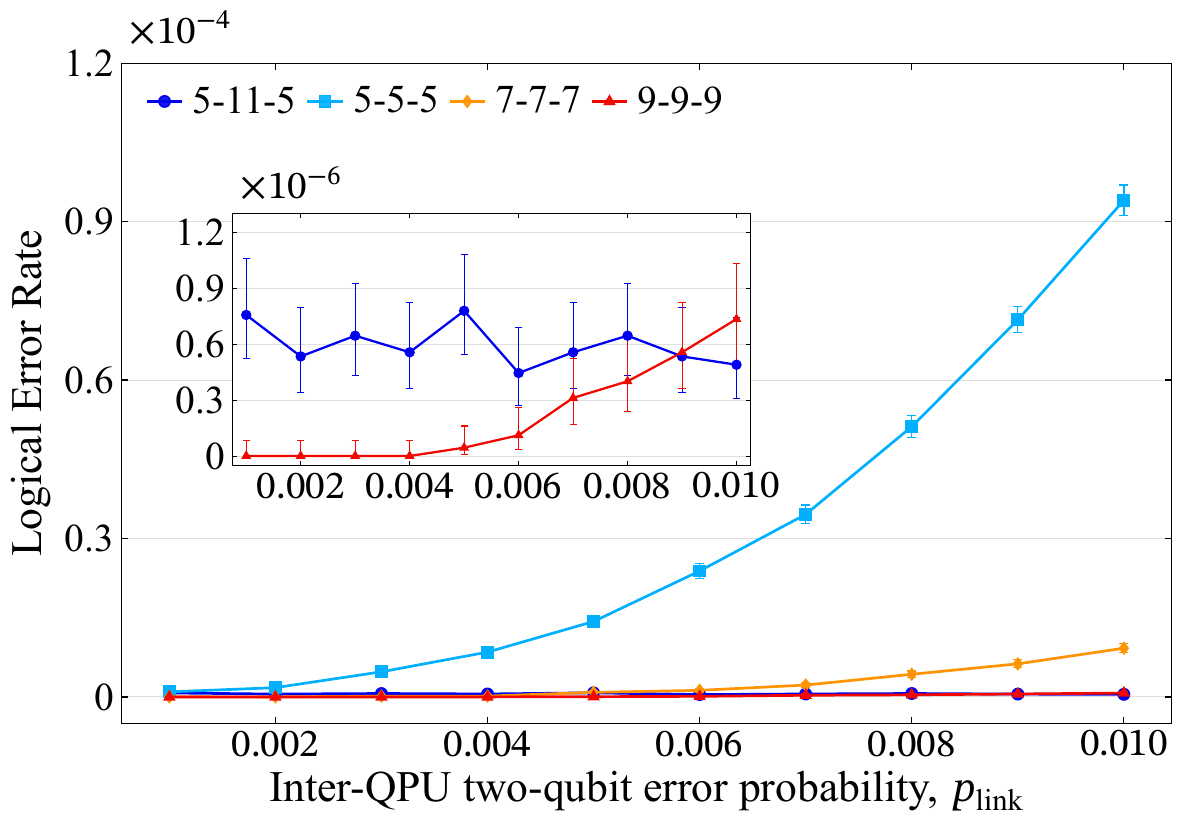}
    \caption{Target-\(Z\) decoding failure probability per CNOT for mixed $5-11-5$ (blue circles) and uniform distance-5, -7, and -9 circuits (cyan squares, orange diamonds, and red triangles), with local noise parameter \(10^{-4}\). The inset resolves the mixed and distance-9 results. Each point uses 45 million shots; error bars show 95\% Clopper--Pearson intervals.}% Our mixed circuit maintains weak dependence on link noise while retaining distance-5 data patches.}

    \label{fig:mixed_large_z}
\end{figure}

Figures~\ref{fig:mixed_large_x} and~\ref{fig:mixed_large_z} illustrate that the larger footprint $5-11-5$ mixed circuit exhibits a similar behavior. Failure probabilities are weakly correlated with \(p_{\mathrm{link}}\), remaining relatively constant for all plotted noise values. At \(p_{\mathrm{link}}=0.01\), failure probabilities for the mixed circuit are comparable to the uniform distance-9 circuit for both control-$X$ and for target-$Z$. In fact, the corresponding $95\%$ confidence intervals overlap, all while our mixed circuit retains distance-5 data patches.

\subsubsection{Observable dependence and traveling direction}

The higher failure probability of the control-$X$ circuit is consistent with the inter-QPU boundary geometry, specifically the orientation of the logical operators relative to the boundary. In the 8-DAM based circuit a conjugate logical $Z$ fault string can intersect the perpendicular logical $X$ operator while following the noisy interface. While the parallel logical $Z$ operator has a beneficial geometric relationship to the interface faults. This orthogonality can produce unequal readout failure probabilities even under strictly unbiased depolarizing noise. This orientation dependence is a known property of surface codes with noisy seams and is expected~\cite{Sinclair2023}. The comparison also depends on the implemented schedules: the $3-7-3$ control-$X$ and target-$Z$ experiments use different joint-parity orders. Their difference therefore cannot be attributed to operator orientation alone. 

We also note that the direction of stabilizer travel is another scheduling consideration. Changing which interactions cross the QPU boundary can change exposure to link faults; simply reordering an unchanged set of gates leaves the number of crossing-gate applications unchanged. Our comparisons evaluate ancilla enlargement together with the implemented syndrome schedules and do not isolate their individual contributions.

\subsubsection{Physical resource requirement}

We quantify the resources used by each simulated circuit by counting distinct physical qubit sites and distinct inter-QPU coupling pairs. The site count includes data, syndrome, and temporary qubits used anywhere in the circuit; it is not a count of peak simultaneous occupancy. Our design requires fewer physical qubits and fewer distinct inter-QPU coupling pairs than the corresponding uniformly enlarged circuits. The $3-7-3$ circuit uses 153 physical qubits and seven crossing pairs, compared with 169 physical qubits and nine pairs for uniform distance-5. The 5-11-5 circuit uses 373 physical qubits and 11 crossing pairs, compared with 521 physical qubits and 17 pairs for uniform distance-9. These correspond to reductions in physical circuit sites of \(9.5\%\) and \(28.4\%\), respectively. 

Distinct coupling pairs specify a connectivity requirement, whereas link-gate applications count repeated uses of those couplings. The $3-7-3$ circuit executes 70 inter-QPU CNOTs, compared with 45 for uniform distance-5; the $5-11-5$ circuit executes 264, compared with 153 for uniform distance-9. The heterogeneous-distance constructions therefore use fewer physical sites and distinct links, but more link-gate applications. Their high-link-noise readout performance is obtained despite this greater number of noisy inter-QPU operations.

\subsubsection{\label{sec:dual_lcnot_data}Simultaneous dual lCNOTs}

\begin{figure*}
    \centering
    \includegraphics[width=\linewidth]{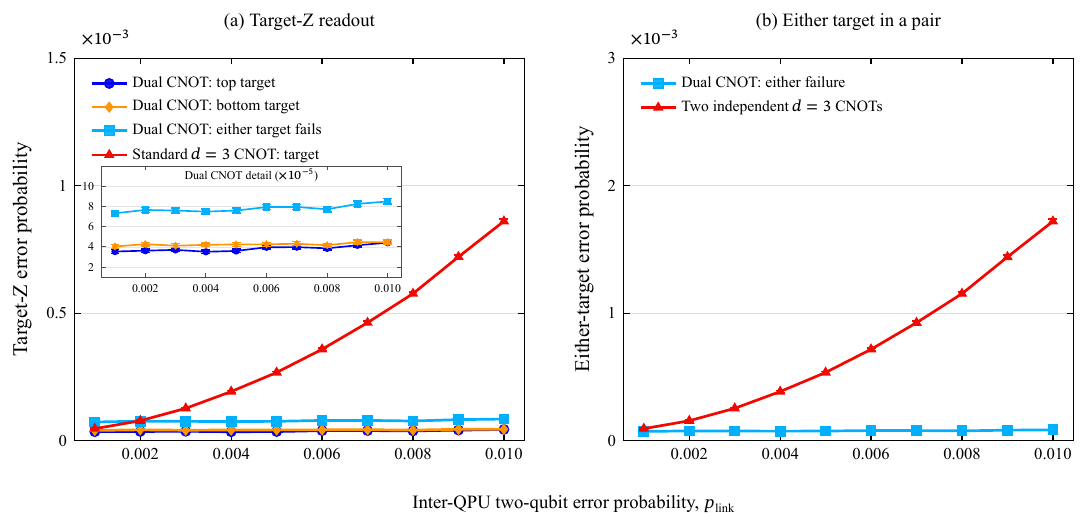}
    \caption{Logical $Z$-readout error rates for the dual-lCNOT circuit and standard $d=3$ lCNOTs. \textbf{(a)} Error rates for the top-target (blue circles), bottom-target (yellow diamonds), and if either of the two lCNOTs fail.  \textbf{(b)} Either-target failure probability compared with two independent standard lCNOTs, calculated as $1-(1-q)^2$ from the single-lCNOT error rate $q$. Error bars indicate 95\% confidence intervals. }

    \label{fig:mixed_dual_z}
\end{figure*}

Figure~\ref{Fig:Dual_lCNOTs} shows our extension of the central ancilla arrangement to four distance-3 data patches, corresponding to lCNOT pairs $(C_1, T_1)$ and $(C_2, T_2)$. Using this geometry and the same simulational model described in Subsection~\ref{sec:sim_model_protocol} we produced the probability failure rate graph shown in Figure~\ref{fig:mixed_dual_z}. Figure~\ref{fig:mixed_dual_z} used a syndrome extraction schedule of six alternating rounds in the first control--ancilla $M_{ZZ}$ merge before the seam qubits are measured out. 

Figure~\ref{fig:mixed_dual_z}(a) shows the individual top- and bottom-target failure probabilities as well as the probability that either target may fail. This joint probability is the measured union found simply by $P_{\text{either}} = P_{\text{top}} + P_{\text{bottom}} - P_{\text{both}}$. Figure~\ref{fig:mixed_dual_z}(b) shows a comparison with two identical independent lCNOT executions using uniform distance-3 code patches. If each has target-$Z$ failure probability $q$, their either-target failure probability is found by $1 - (1 - q)^2$. Neither top- or bottom-targets in the mixed geometry simulations exhibit the pronounced link-noise growth seen in the standard reference.

Taken together, the results motivate concentrating additional code resources at noisy inter-QPU interfaces while retaining compact data patches. The single-lCNOT comparisons demonstrate a tradeoff relevant to modular architectures with limited physical-qubit capacity or interface connectivity: fewer physical sites and distinct couplings in exchange
for more repeated link operations. Assessing the resulting hardware throughput requires link-generation times, syndrome-round durations, and idle errors, which are outside the present model.

\section{\label{sec:conclusion}Conclusion and Outlook}

The 8-DAM architecture is well suited for distributed and modular quantum error correction codes, particularly lattice-surgery. By concentrating additional error-correction code resources at a noisy inter-QPU interface while retaining more compact data patches, 8-DAM, combined with a traveling stabilizer syndrome extraction schedule produces failure probabilities that depend less strongly on link noise than in an architecture that uses uniform-distance circuits. 

We have presented simulated $3-7-3$ and $5-11-5$ circuits using 8-DAM that, at the upper end of EFT-era physical link-noise probabilities, are able to achieve comparable to or better performance than uniformly enlarged patches. Additionally, we have shown that 8-DAM is capable of implementing a shared-ancilla geometry to support the simultaneous execution of dual-CNOTs.

Our study has found an illuminating tradeoff between physical resources and the repeated use of a quantum interconnect. The
$3-7-3$ and $5-11-5$ circuits use $9.5\%$ and
$28.4\%$ fewer distinct physical circuit sites than the uniform
distance-5 and distance-9 comparisons, respectively, together
with fewer distinct inter-QPU coupling pairs. Their schedules,
however, require more inter-QPU gate applications over the
complete circuit. Selective ancilla enlargement is therefore
particularly relevant when physical-qubit capacity or interface
connectivity is constrained, with its practical benefit also
depending on the cost of repeatedly using each link.

For the boundary modifications examined, we observed the same
qualitative error-suppression behavior. A broader comparison
of interface geometries remains an important direction,
including further development of the $45^\circ$-rotated
design. Jointly optimizing the patch layout, boundary checks,
and traveling stabilizer schedule may reduce the number of
inter-QPU gate applications while preserving local operations
within each processor and the required circuit fault distance.
Controlled comparisons would also help distinguish the
contributions of ancilla enlargement and syndrome scheduling
to the observed performance.

Our simulations characterize selected logical-readout failure
probabilities under a circuit-level noise model. Extending
this analysis to include link-generation failures and latency,
idle errors, and correlated faults would clarify the hardware
conditions under which the proposed resource tradeoff is
advantageous. These extensions, together with comparisons
of alternative layouts and concurrent operation schedules,
will help determine how selective protection of noisy
interfaces can improve the performance of modular quantum
computers.
\section{\label{sec:code_avail}Code availability}

The Stim circuits and the codes used for the simulations in this research can be found on our \hyperlink{https://github.com/aaj10789/8DAM/}{GitHub}~\cite{Jebraeilli_Dilly}.

\begin{acknowledgments}
This material is based upon work supported by the U.S. Department of Energy, Office of Science,Advanced Scientific Computing Research (ASCR) program as part of the Distributed Quantum Computing Algorithms Lab LDRD.
\par\medskip
\begingroup
\small
\noindent\fbox{%
\parbox[b]{\dimexpr\linewidth-2\fboxsep-2\fboxrule\relax}{%
    The submitted manuscript has been created by UChicago Argonne, LLC, Operator of 
    Argonne National Laboratory (``Argonne''). Argonne, a U.S.\ Department of 
    Energy Office of Science laboratory, is operated under Contract No.\ 
    DE-AC02-06CH11357. 
    The U.S.\ Government retains for itself, and others acting on its behalf, a 
    paid-up nonexclusive, irrevocable worldwide license in said article to 
    reproduce, prepare derivative works, distribute copies to the public, and 
    perform publicly and display publicly, by or on behalf of the Government.  The 
    Department of Energy will provide public access to these results of federally 
    sponsored research in accordance with the DOE Public Access Plan. 
    \url{http://energy.gov/downloads/doe-public-access-plan}.
  }%
}
\par
\endgroup
\end{acknowledgments}

\appendix
\nocite{}
\bibliography{bibliography}

\end{document}